\documentstyle[11pt]{article}

\title{A TOPOLOGICAL MODEL FOR TWO-DIMENSIONAL GRAVITY COUPLED TO MATTER}
\author{L.F.Cugliandolo \thanks{Permanent address: Departamento de
F\'{\i}sica, Universidad Nacional de La Plata, Argentina; CONICET, Argentina}
\\
International Centre for Theoretical Physics, Trieste, Italy \\
F.A.Schaposnik\thanks{Investigador CIC-BA, Argentina} and
H.Vucetich$^*$ \\
Departamento de F\'{\i}sica, Universidad
Nacional de La Plata \thanks{Postal address: CC67 (1900) La Plata, Argentina}}
\date{June 1991}
\begin{document}
\maketitle

\newpage

\begin{abstract}

Starting from a topological gauge theory in two dimensions with symmetry
groups $ISO(2,1)$, $SO(2,1)$ and $SO(1,2)$ we construct a model for gravity
with non-trivial coupling to matter. We discuss the equations of motion
which are connected to those of previous related models but incorporate
matter content. We
also discuss the resulting quantum theory and finally present explicit
formul\ae $\;$ for topological invariants.

\end{abstract}

\newpage
\section{INTRODUCTION}

Lower dimensional theories of gravity have recently attracted much attention
\cite{JT}-\cite{CH}.
In particular, considerable progress has been achieved by exploiting the
connection between two and three dimensional gravity models and Topological
Quantum Field Theories (TQFTs)
\cite{W2}-\cite{Blau}.
In this way, it has been proven that general
relativity in three dimensions is equivalent to the Chern-Simons theory (with
gauge groups $ISO(2,1)$, $SO(3,1)$ or $SO(2,2)$ depending on the value of the
cosmological constant).
Using this connection, it has been shown that the
theory is renormalizable and finite, and that it can be solved exactly
\cite{W2}.

Since there is no Chern-Simons like term in two dimensions, other types of
TQFTs have to be employed in an attempt to construct two-dimensional gravity
models starting from gauge theories \cite{CHW}-\cite{Blau}.
It is the purpose of this work to
present one of such models based on a two-dimensional TQFT which not only
includes gauge fields (with symmetry groups $ISO(1,1)$, $SO(2,1)$ or $SO(1,2)$)
but also scalar fields, naturally leading to a description of gravity
coupled to matter.

One way in which topological theories (of the Witten
type \cite{W1}) can be obtained is by quantizing a classical action
$S_{cl}$ that corresponds to a topological invariant \cite{BS}. In a sense,
these
classical actions are trivial since, being by essence invariant under
arbitrary transformations, all fields can be gauged away at the classical
level. At the quantum level, this reflects in the reduction of the solution
space from an infinite dimensional to a finite dimensional one. The
resulting quantum action is intimately related to instanton configurations
carrying the topological charge. In order to have instantons in
two-dimensional gauge theories, one necessarily has to add Higgs fields. In
the Abelian case these instantons are the time-honoured Nielsen-Olesen
vortices carrying a topological charge $Q\in Z$ related to the vortex magnetic
flux \cite{Nielsen}. Non-Abelian extensions can be constructed and the
resulting
instantons
are again vortex-like configurations. The topological charge is
again associated to the magnetic flux and takes the form \cite{Fidel}:

\begin{equation}
Q = \frac{1}{2\pi}\int d^2x \sqrt g E^{\mu\nu} < \Psi F_{\mu\nu}>
\; .
\label{1}
\end{equation}
Here $\Psi$ is one of the scalar fields in the adjoint representation of
the gauge group, $F_{\mu\nu}$ is the gauge field curvature and $``<\;>"$
represents an adequate inner product.

Starting from an action of the form (\ref{1}) and identifying the gauge field
with Zweibein and spin connection fields, a highly non-trivial model for
two-dimensional gravity has been constructed by Chamseddine and Wyler
\cite{CHW} (see also \cite{MS}-\cite{Blau}).
In this model, there is a scalar field which just plays the r\^ole of
a Lagrange multiplier. Any attempt to add to this topological action kinetic
energy terms for $\Psi$ either breaks the covariance of the model or implies
the appearence of rather complicated self-interactions which obscure the
resulting theory.

There is another possibility for constructing TQFTs put forward by
Labastida and Pernici \cite{LP} (see also \cite{BRT}).
In their approach, instead of starting from a $S_{cl}$
which is a topological charge, one constructs a gaussian action in which
instanton defining equations (Bogomol'nyi equations in the two-dimensional
case) have a relevant r\^ole. All fields enter in this action in a self-dual
way (in a sense to be precised in next sections) and hence kinetic terms
for scalars appear naturally in a way that does not imply a metric dependence
at the quantum level. It is this approach
the one we follow in the present work. We think it is the most natural one,
especially if one takes into account the central r\^ole that instanton moduli
space plays in TQFTs and the fact that in two-dimensional gauge theories
instanton equations have non-trivial solutions only when an appropriate
number of scalar fields, with their corresponding kinetic terms, are
included.

In this way, we arrive at a model for two-dimensional gravity
non-trivially coupled to matter. After reviewing the two-dimensional
topological
gauge theory in Section 2, we establish the connection with the gravity
model in Section 3. We there identify Zweibein and spin connection fields
and discuss the resulting classical equations of motion. These equations
reduce to the Jackiw-Teitelboim equations \cite{JT} when scalars are absent and
also include as a particular case Chamseddine-Wyler ones \cite{CHW}. In
Section 4 we discuss
the quantum action and its symmetries leaving for Section 5 the evaluation
of topological invariants. Finally we present a discussion of the model
and the conclusions to our work in Section 6.

\section{THE GAUGE MODEL}
\setcounter{equation}{0}
\label{Section2}

In this Section we briefly review the non-Abelian two-dimensional
topological field theory
constructed in ref.\cite{CLS1}, which is at the basis of the model for
two-dimensional gravity to be presented in Section \ref{Section3}.

The model for a non-Abelian two-dimensional gauge field theory that we
consider has been
constructed, in the manner of Labastida-Pernici \cite{LP}, starting
from a classical action defined on a general two-dimensional manifold $M$, in
which Bogomol'nyi equations play a central r\^ole,
\begin{equation}
S_{cl}[M]= \int_M d^2x \sqrt{g} < (H_{\mu\nu} - B_{\mu\nu})^2
+|H_\mu - B_\mu|^2 + |\tilde  H_\mu - \tilde  B_\mu|^2 > .
\label{2.1}
\end{equation}
In this expression, $H_{\mu\nu}$, $H_\mu$ and $\tilde  H_\mu$ are auxiliary
fields belonging to the algebra of the gauge group ${\cal G}$; $H_{\mu\nu}=
H_{\mu\nu}^A T_A$,
$H_\mu=H_\mu^A T_A$ and $\tilde  H_\mu=\tilde  H_\mu^A T_A$, where $T_A$,
$A=1,..., dim {\cal G}$, are the group generators. They are self-dual fields in
the sense that \cite{CLS1}
\begin{equation}
H^{\mu\nu}=E^{\mu\nu} H
\label{2.2}
\end{equation}
with $H$ the dual of $H^{\mu\nu}$ and ${^*}H^\mu$ the dual of $H_\mu$
satisfying
\begin{equation}
^*H^\mu \equiv i E^{\mu\nu} H_\nu = H^\mu
\label{2.3}
\end{equation}
($E^{\mu\nu}$ is the contravariant two-dimensional Levi-Civita tensor,
$E^{\mu\nu}=\frac{\epsilon^{\mu\nu}}{\sqrt{g}}$,
$\epsilon^{01}=-\epsilon^{10}=1$). $B_{\mu\nu}$, $B_\mu$ and $\tilde  B_\mu$
stand for the following expressions
\begin{eqnarray}
B_{\mu\nu}      &\equiv& F_{\mu\nu} - e_G E_{\mu\nu} \Psi < \Phi^2 -\Phi_0^2 >
\; , \label{2.4} \\
B_\mu           &\equiv& D_\mu^+ \Phi + E^+_{\mu\nu} \; [ \Psi, D^\nu  \Phi ]
\; ,  \label{2.5} \\
\tilde  B_\mu &\equiv& D_\mu^+ \Psi
\; ,
\label{2.6}
\end{eqnarray}
so that
\begin{eqnarray}
B_{\mu\nu}      &=& 0
\; , \label{2.7} \\
B_\mu           &=& 0
\; , \label{2.8} \\
\tilde  B_\mu &=& 0
\label{2.9}
\end{eqnarray}
represent the Bogomol'nyi equations corresponding to a two-dimensional
non-Abelian gauge theory \cite{CLS2}.
These equations have to be supplemented
with the constraints
\begin{eqnarray}
<\Psi^2> &=& 1
\; , \label{2.10} \\
<\Psi \Phi> &=& 0
\; .
\label{2.11}
\end{eqnarray}
(see ref.\cite{CLS2} for details).
The dynamical fields of the theory defined by $S_{cl}$ are,
then,
a gauge field $A_\mu$ taking values in the algebra of the gauge group
${\cal G}$, $A_\mu= A_\mu^A T_A$, and two scalar fields  $\Psi$ and $\Phi$
in the adjoint representation of the group ${\cal G}$, $\Psi=\psi^A T_A$
and $\Phi= \phi^A T_A$. The field strength $F_{\mu\nu}$ is defined as
\begin{equation}
F_{\mu\nu}= F_{\mu\nu}^A T_A \equiv \partial_\mu A_\nu -\partial_\nu A_\mu
+ e_G [ A_\mu, A_\nu]
\; ,
\end{equation}
accordingly, the covariant derivative $D_\mu$ is defined as
\begin{equation}
D_\mu \equiv \partial_\mu + e_G [A_\mu, \;]
\; .
\label{2.13}
\end{equation}

The ``$^+$'' symbol appearing in covariant derivatives of expressions
(\ref{2.5})
and (\ref{2.6}) is defined as follows. Given a vector $C_\mu$,
\begin{equation}
C_\mu^+ \equiv \frac{1}{2} (C_\mu + ^* C_\mu) = \frac{1}{2}
(C_\mu + i E_{\mu\nu} C^\nu)
\; .
\label{2.14}
\end{equation}
It is easy to prove that $C_\mu^+$ is a self-dual vector (cf. eq.(\ref{2.3}));
this property
implies the self-duality of the expressions (\ref{2.5}) and (\ref{2.6})
and, furthermore, as we shall see, {\it the independence of the quantum action
on the metric $g_{\mu\nu}$}. This quality is at the root of the
topological character of the quantum theory constructed from $S_{cl}$
\cite{W1}; it means that the partition function constructed from $S_{cl}$
does not depend on the choice of any particular metric, it can only depend
on the topology of $M$.

The ``$<\;>$'' symbol denotes the appropriate invariant, non-degenerate and
associative inner product. Its explicit definition depends on the choice
of the group ${\cal G}$ and will be presented, in the cases of interest,
in the next Section.

Bogomol'nyi equations (\ref{2.7})-(\ref{2.9}) are first order differential
equations whose solutions solve the (second order) Euler-Lagrange equations
for a non-Abelian gauge theory coupled to two Higgs fields defined on the two
dimensional manifold $M$ \cite{CLS2}. The number of Higgs
fields
introduced is such that complete symmetry breaking is achieved so as to
ensure non-trivial topology for the gauge field $A_\mu$.
In general, Bogomol'nyi
equations exist whenever a particular relation
between coupling constants hold. For instance, for the model to be considered,
two scalar fields are necessary and a potential ensuring complete symmetry
breaking is
\begin{equation}
V(\Psi, \Phi)= g_1 <(\Phi^2-\Phi^2_0)^2>  + g_2  <(\Psi^2 -1)^2> + g_3 <(\Psi
\Phi)^2>
\label{2.15}
\end{equation}
but, due to conditions (\ref{2.10}) and (\ref{2.11}), only the first term
plays a r\^ole in the model.
In this case, Bogomol'nyi condition relates the gauge coupling constant
$e_G$ and the potential strenght $g_1$,
\begin{equation}
e_G^2= k g_1
\; ,
\label{2.16}
\end{equation}
the numerical constant $k$ being determined by the inner product definition.

The peculiarity of an action like (\ref{2.1}) is its invariance under
the most general transformations of the dynamical fields (in this case $A_\mu$,
$\Psi$ and $\Phi$), provided one adequately choose the transformation laws
for the auxiliary fields (namely $H_{\mu\nu}$, $H_\mu$ and $\tilde  H_\mu$).
Indeed, if one transforms the gauge and scalar fields in the most general
form
\begin{eqnarray}
\delta A_\mu &=& \epsilon_\mu - D_\mu \epsilon
\; , \label{2.17} \\
\delta \Phi &=& \theta - e_G [\Phi, \epsilon]
\; , \label{2.18} \\
\delta \Psi &=& \tilde  \theta - e_G [\Psi, \epsilon]
\label{2.19}
\end{eqnarray}
(we have distinguished usual gauge transformations for later convenience),
the classical action (\ref{2.1}) remains unchanged provided
\begin{eqnarray}
\delta H_{\mu\nu} &=& \delta B_{\mu\nu} + [ H_{\mu\nu} - B_{\mu\nu}, \epsilon]
\; , \label{2.20} \\
\delta H_\mu &=& \delta B_\mu + [H_\mu - B_\mu, \epsilon]
\; , \label{2.21} \\
\delta \tilde  H_\mu &=& \delta \tilde  B_\mu + [\tilde  H_\mu -
\tilde  B_\mu, \epsilon]
\; , \label{2.22}
\end{eqnarray}
where variations in the right hand side are to be computed in terms of
the variations of the dynamical fields (\ref{2.17})-(\ref{2.19}). It is
important to note that not
all of the parameters are effective regarding these transformations: if one
chooses $\epsilon_\mu=D_\mu \xi$, $\epsilon=\xi$, $\theta=e_G [\Phi,\xi]$ and
$\tilde\theta=e_G [\Psi,\xi]$ not only the action remains invariant but also
all fields do it (on shell).

Transformations (\ref{2.17})-(\ref{2.19}) are enough
to select a gauge in which $H_{\mu\nu}$, $H_\mu$ and $\tilde  H_\mu$ vanish.
This can be achieved with parameters $\epsilon_\mu$, $\theta$ and $\tilde
\theta$ and leaving untouched parameter $\epsilon$ \cite{CLS1}. In this
gauge, the equations of motion of $S_{cl}$ coincide
with the Bogomol'nyi equations (\ref{2.7})-(\ref{2.9}).

This symmetry of the classical action is called a ``large'' or
``topological'' symmetry and has to be fixed in the process of quantization.
The second generation gauge
invariance mentioned in a previous paragraph imposes a refined
BRST quantization, for instance by using Batalin-Vilkovisky method \cite{BV}.
This has been done in detail in ref.\cite{CLS1}. Just for completeness let
us indicate the main lines of the quantization procedure. A generic term
of the classical action (\ref{2.1}) takes the form
\begin{equation}
S_{cl}^1 =  \int_M d^2 x \sqrt g  <|H- B[\Phi]|^2 >
\; ,
\label{S1}
\end{equation}
where $\Phi$ is a collection of dynamical fields, $B[\Phi]=0$ is the associated
Bogomol'nyi equation and $H$ is the corresponding auxiliary field. Evidently,
each one of the terms in (\ref{2.1}) has this form. This action remains
invariant under the large transformations
\begin{eqnarray}
\Phi &\rightarrow& \Phi + \delta \Phi \; ,  \label{inv1}  \\
H &\rightarrow& H + \frac{\delta B}{\delta \Phi} \delta \Phi
\; . \label{inv2}
\end{eqnarray}
Associated with transformations (\ref{inv1})-(\ref{inv2}), we can define BRST
commutators
\begin{eqnarray}
\{Q, \Phi \} &=& \chi \; , \\
\{Q, H \} &=& \frac{\delta B}{\delta \Phi} \chi \; .
\end{eqnarray}
The linear transformation $\{Q,\; \}$
is defined by stating that the BRST
transformation of a functional $F$ is
\begin{equation}
\delta^{BRST} F= \lambda \{Q, F \}
\; ,
\label{2.24}
\end{equation}
with $\lambda$ a Grassmann odd constant parameter.
$\chi$ represents the ghost related to the symmetry (\ref{inv1})-(\ref{inv2}).
Proceeding \`a la Batalin-Vilkovisky,
the quantum action is constructed from $S_{cl}$ as follows
\begin{equation}
S_q^1= S_{cl} +\{Q, F\}
\label{F}
\end{equation}
where $F$ is a ``gauge fermion'' \cite{BV} introduced to fix the large
symmetry. It
is evident from (\ref{inv2}) that $H$ can be gauged away; to this end, one
chooses
\begin{equation}
F= \int_M d^2x \sqrt g < X H >
\; ,
\end{equation}
being $X$ an antighost field. We impose the following BRST
transformation laws on the antighost field $X$ and Lagrange multiplier $D$
\begin{eqnarray}
\begin{array}{rcl}
\{Q, X\} &=& D \; , \\
\{Q, D\}&=& 0 \; ,
\end{array}
\end{eqnarray}
with this $S_q^1$ becomes
\begin{equation}
S_q^1 = S_{cl}^1 +  \int_M \; d^2x
\sqrt g <X \frac{\delta B}{\delta \Phi} \chi + DH >
\; .
\label{SQ1}
\end{equation}
The second term in (\ref{SQ1}) corresponds to the ghost action and the third
one to a gauge fixing action.
Thus, the partition function ${\cal Z}$ for a classical action of the kind
(\ref{S1}) with a large symmetry (\ref{inv1})-(\ref{inv2}) is
\begin{eqnarray}
{\cal Z}&=&\int {\cal D} \Phi \;  {\cal D} H  \;  {\cal D} \chi  \;  {\cal D}
X  \;  {\cal D} D \;  \nonumber \\
& & \hspace{0.6in} \exp [- \int_M d^2 x \sqrt g < |H-B|^2 +
X \frac{\delta B}{\delta \Phi} \chi + DH >] \nonumber \\
&=& \int  {\cal D} \Phi \;  {\cal D} \chi  \;  {\cal D} X
\;  \exp[- \int_M d^2 x \sqrt g < B^2 + X \frac{\delta
B}{\delta \Phi} \chi >] \;.
\end{eqnarray}

The quantization of the classical action under consideration,
eq.(\ref{2.1}), follows the same steps. However, the actual transformations
(\ref{2.17})-(\ref{2.22}) are slightly more complicated and hence our arguments
have to be generalized to also include, appart from ghosts of the
$\chi$-type associated with the large symmetry, ghosts associated with the
usual
 gauge
and second generation invariances. In any case, the final form of $S_q$ can be
written in the form
(\ref{F}) where $F$ is some functional of the original fields (gauge, scalar
and auxiliary fields) and new fields (ghosts and Lagrange multipliers)
introduced in the gauge fixing procedure. Moreover, it can be shown that there
exists a functional $V$ such that
$S_q$ can be written as a BRST commutator
\begin{equation}
S_q=\{Q, V\}
\; . \label{2.25}
\end{equation}
{}From this equation it is easy to show that
\begin{equation}
\frac{\delta {\cal Z}}{\delta
g_{\mu\nu}}=0
\; ,
\label{Z}
\end{equation}
the defining equation for TQFTs. Furthermore, the partition function is, for
similar reasons, independent
of the gauge coupling constant $e_G$, as long as $e_G$ is different from zero.
This can be easily demonstrated by going through new field variables
in such a way that $\frac{1}{e_G^2}$ is factorized from the quantum action,
which remains gauge coupling independent.
This property
permits to exactly evaluate ${\cal Z}$ by going to the small
$e_G^2$ limit where it is dominated by the classical minima, that is, the
solutions to Bogomol'nyi equations.
In the next Section we shall give an explicit form for $S_q$ \cite{CLS1}.

\section{THE GRAVITATIONAL MODEL}
\label{Section3}
\setcounter{equation}{0}

Let us now construct a model for two-dimensional gravity based
on the topological model presented in Section \ref{Section2}. We consider
the symmetry groups $ISO(1,1)$, $SO(2,1)$ and $SO(1,2)$, with generators
which will be identified with the $T_A$'s introduced in the previous Section.
In order to describe the three cases with a sole algebra, we write
$(T_A)=(P_a, J)$, $a=0,1$,
\begin{eqnarray}
{[}P_a,P_b{]} &=& \Lambda \epsilon_{ab} J \; , \label{3.1} \nonumber\\
{[}P_a,J{]} &=& -\epsilon_a^{\, b} P_b \; , \label{3.2} \nonumber \\
{[}J,J{]} &=& 0
\; . \label{3.3} \nonumber
\end{eqnarray}
In this context, $P_a$ and $J$ will play the r\^ole of the generators of the
two translations and Lorentz rotation, respectively, on the two-dimensional
manifold $M$. Latin indices are raised
and lowered with an internal metric $\eta_{ab}$. We shall see that a choice
of signature for the $\eta_{ab}$ will fix the corresponding signature for
the metric in our gravity model. With our conventions $\epsilon^{ab}$
is such that $\epsilon^{01}=-\epsilon^{10}=1$.

The constant $\Lambda$ behaves as a dimensionless cosmological
constant\footnote{Dimensionfull magnitudes should be constructed by using the
gauge coupling constant $e_G$ which has dimensions of mass.}; the
values $\Lambda=0$, $\Lambda>0$ and $\Lambda<0$ give rise to the $ISO(1,1)$
group (the isometry group of two-dimensional flat Minkowski space-time), the
$SO(2,1)$ group
(the isometry group of two-dimensional de Sitter space-times) and the
$SO(1,2)$ group (the isometry group of two-dimensional anti-de Sitter
space-times), respectively.

For the $SO(2,1)$ and $SO(1,2)$ groups we define the inner product by
using the Killing metric arising from their algebras; thus,
\begin{eqnarray}
<P_a, P_b> &=& \Lambda \eta_{ab}
\; , \label{3.4}  \nonumber\\
<P_a, J> &=& 0
\; , \label{3.5}  \nonumber\\
<J,J> &=& 1
\; . \label{3.6}
\end{eqnarray}
We cannot proceed in an analogous way in the case of the $ISO(1,1)$ group
because
of the degeneracy of its Killing metric. We can, however, overcome this
difficulty by defining the following inner product \cite{MS}
\begin{eqnarray}
<P_a, P_b> &=& \eta_{ab}
\; , \label{3.7} \\
<P_a, J> &=& 0
\; , \label{3.8} \\
<J,J> &=& 1
\; . \label{3.9}
\end{eqnarray}
It is not possible, though, to avoid the degeneracy of the Casimir operator
which is still taken in the form $W=P^a P_a$.

Since our aim is to make contact with two-dimensional gravity, we introduce
the following notation
\begin{eqnarray}
A_\mu^a&=&e_\mu^a
\; , \label{3.10} \\
A_\mu^3 &=& f_\mu
\; , \label{3.11}
\end{eqnarray}
attempting to identify two of the vector potential components (those along the
``translation directions'' $P_a$) as a Zweibein and to relate the remaining
vector potential component
(the one along the ``Lorentz rotation'' direction $J$) to the spin connection.
Then, the covariant derivative $D_\mu$ (eq.(\ref{2.13})) acting on an algebra
valued field $C=(c^a, c)$ becomes
\begin{equation}
D_\mu C = D_\mu [e,f] C = {\cal D}^{ab}_\mu [f] c_b P_a +
(\partial_\mu c + e_G \Lambda \; \epsilon_{ab} e_\mu^a c^b ) J
\; ,
\label{cov}
\end{equation}
where ${\cal D}_\mu^{ab} [f]$ is given by
\begin{equation}
{\cal D}_\mu^{ab} [f] \equiv \delta^{ab} \partial_\mu - e_G \epsilon^{ab} f_\mu
\; .
\label{3.22'}
\end{equation}
Concerning the Higgs fields, we denote them
\begin{eqnarray}
\Psi &=& (\psi^a, \psi)
\; , \label{3.12} \\
\Phi &=& (\phi^a, \phi)
\; , \label{3.13}
\end{eqnarray}
while for auxiliary fields $H_{\mu\nu}$, $H_\mu$ and $\tilde H_\mu$ and
expressions $B_{\mu\nu}$, $B_\mu$ and $\tilde B_\mu$ we write
\begin{eqnarray}
\begin{array}{rclrcl}
H_{\mu\nu} &=& (h_{\mu\nu}^a, h_{\mu\nu}) &
B_{\mu\nu} &=& (b_{\mu\nu}^a, b_{\mu\nu}) \; , \\
H_{\mu} &=& (h_\mu^a, h_\mu) &
B_\mu &=& (b_\mu^a, b_\mu) \; , \\
\tilde H_\mu &=& (\tilde h_\mu^a, \tilde h_\mu) &
\tilde B_\mu &=& (\tilde b_\mu^a, \tilde b_\mu) \; .
\end{array}
\end{eqnarray}
Then,
\begin{eqnarray}
b_{\mu\nu}^a &=& {\cal D}_\mu^{ab} {[}f{]} e_{\nu b} -
{\cal D}_\nu^{ab} {[}f{]} e_{\mu b}- e_G E_{\mu\nu} \, \psi^a \, v(\Phi) \; ,
\label{b1} \\
b_{\mu\nu} &=& \partial_\mu f_\nu - \partial_\nu f_\mu +  e_G \Lambda
E_{\mu\nu} - e_G E_{\mu\nu} \, \psi \, v(\Phi)
\; , \\
b_\mu^a &=& \Delta_{\mu}^{+a} (f,\phi_c,\phi)
- E_{\mu\nu}^+ \epsilon^{ab} \psi
\Delta^\nu_b (f,\phi_c,\phi) + \nonumber \\
& & E_{\mu\nu}^+ \epsilon^{ab} \, \psi_b \, \partial^\nu \phi
\; , \\
b_\mu &=& \partial_\mu^+ \phi + e_G \Lambda \, \epsilon_{ab} \, e_\mu^{+a}
\psi^b
+ \Lambda \, E_{\mu\nu}^+ \epsilon^{ab} \, \psi_a \,
\Delta^\nu_b(f,\phi_c,\phi)
\; , \\
\tilde b_\mu^a &=& \Delta_\mu^{+a}(f,\psi_b,\psi)
\; ,  \\
\tilde b_\mu &=& \partial_\mu^+ \psi + e_G \Lambda \, \epsilon_{ab} \,
e_\mu^{+a} \, \psi^b
\; .
\label{b2}
\end{eqnarray}
where
\begin{equation}
v(\Phi) \equiv  <\Phi^2 - \Phi_0^2>
\; ,
\end{equation}
and $\Delta_\mu^a(f,\phi_b,\phi)$ stands for
\begin{equation}
\Delta_\mu^a(f,\phi_b,\phi) \equiv {\cal D}_\mu^{ab} {[}f{]} \phi_b
+e_G \epsilon^{ab} \, e_{\mu b} \, \phi
\; .
\end{equation}
With this notation, the equations of motion of the theory
become
\begin{eqnarray}
h_{\mu\nu}^a &=& b_{\mu\nu}^a  \; , \label{3.14} \\
h_{\mu\nu} &=& b_{\mu\nu} \; , \label{3.15} \\
h_\mu^a &=& b_\mu^a \; , \label{3.16} \\
h_\mu &=& b_\mu \; , \label{3.17} \\
\tilde h_\mu^a &=& \tilde b_\mu^a \label{3.18} \\
\tilde h_\mu &=& \tilde b_\mu  \; . \label{3.19}
\end{eqnarray}
Similarly, the constraints ({\ref{2.10}) and (\ref{2.11}) are
\begin{eqnarray}
\Lambda \psi^a \psi_a + \psi^2 &=& 1
\; , \label{3.20} \\
\Lambda \psi^a \phi_a + \psi \phi &=& 0
\; , \label{3.21}
\end{eqnarray}
in the cases of the $SO(2,1)$ and $SO(1,2)$ groups and
\begin{eqnarray}
\psi^a \psi_a + \psi^2 &=& 1
\; , \label{3.22} \\
\psi^a \phi_a + \psi \phi &=& 0
\; ,  \label{3.23}
\end{eqnarray}
in the case of the $ISO(1,1)$ group.

In order to confirm the identification between the gauge field components
$A_\mu^a$
and the Zweibein $e_\mu^a$ (so as to interpret the topological model presented
in Section \ref{Section2} as a model for two-dimensional gravity) it is
convenient at this point to analyse Bogomol'nyi equations which are, in fact,
the equations
of motion for the topological model with quantum action $S_q$ (\ref{2.25})
in the small $e_G$ limit. As we stated above, the gauge freedom (see
eqs.(\ref{2.17})-(\ref{2.22})) allows us to gauge away auxiliary fields
$H_{\mu\nu}$, $H_\mu$ and $\tilde H_\mu$ so that the equations of motion
(\ref{3.14})-(\ref{3.19}) become the Bogomol'nyi equations.
Furthermore, as we explained at the end of
the previous Section, ${\cal Z}$ is independent of $e_G$ and can be evaluated
by taking the limit for which the path integral is dominated
by Bogomol'nyi equations solutions.

As we shall show below, the first two equations of our gravity model
become equations for torsion and curvature similar to those presented in
ref.\cite{JT} but with extra terms added to the cosmological constant. The
rest of the equations are directly related to the matter content of the
system. To see this, let us obtain from eq.(\ref{3.14}) an explicit expression
for $f_\mu$ in terms of $e_\mu^a$, $\psi^a$ and $\Phi$ under
the assumption that $e_\mu^a$ is invertible (i.e. there exists $e^\mu_b$ such
that, $e_\mu^a e^\mu_b=\delta^a_b$
and $e_\mu^a e^\nu_a=\delta_\mu^\nu$). The expression is
\begin{equation}
f_\mu=\frac{1}{e_G} E^{\alpha\beta} (\partial_\alpha e_\beta^a) e_{\mu a} -
e_\mu^a \psi_a
v(\Phi)
\; . \label{3.222}
\end{equation}
Using it, eq.(\ref{3.15}) transforms into
\begin{equation}
\frac{2}{e_G} E^{\mu\nu} \partial_\mu w_\nu - 2 E^{\mu\nu} \partial_\mu
(e_\nu^a
\psi_a
v(\Phi)) - 2 e_G \, \psi v(\Phi) + 2 e_G \Lambda =0
\label{3.23'}
\end{equation}
where $w_\mu$ is defined as follows
\begin{equation}
w_\mu \equiv E^{\alpha\beta} (\partial_\alpha e_\beta^a) e_{\mu a}
\; . \label{3.24}
\end{equation}
{}From eqs.(\ref{3.23'}) and (\ref{3.24}), one can see that it is consistent to
identify
$e_\mu^a$ with a Zweibein so that the two-dimensional metric $g_{\mu\nu}$
be given by
\begin{equation}
g_{\mu\nu}= e_\mu^a e_\nu^b \eta_{ab}
\; . \label{3.25}
\end{equation}
Indeed, in two dimensions the affine spin connection (under the condition of
metricity) can be written in the form
\begin{equation}
w_\mu^{ab} = \epsilon^{ab} \Omega_\mu
\; .
\label{3.25'}
\end{equation}
If we identify the connection $\Omega_\mu$ with $w_\mu$ as given by
eq.(\ref{3.24}),
\begin{equation}
w_\mu^{ab}= \epsilon^{ab} w_\mu
\; ,
\label{3.26}
\end{equation}
the first
term in the left hand side of eq.(\ref{3.23'}) becomes proportional to the
curvature scalar $R$,
\begin{equation}
R=2 E^{\mu\nu} \partial_\mu w_\nu
\label{3.27}
\end{equation}
and the complete equation of motion (\ref{3.23'}) takes the form
\begin{equation}
R+2 e_G^2 \Lambda = 2 e_G \, E^{\mu\nu} \partial_\mu (e_\nu^a \psi_a v(\Phi)) -
2 e_G^2 \, \psi
v(\Phi) \equiv e_G \tau
\; .
\label{3.28}
\end{equation}
Were the scalar field $\Phi$ absent, this equation would reduce to
\begin{equation}
R+2 e_G^2 \Lambda=0
\; . \label{3.29}
\end{equation}
This is precisely one of the equations of motion for the Jackiw-Teitelboim
model \cite{JT} for two-dimensional gravity
(also discussed
in refs.\cite{CHW}-\cite{IT}). Moreover, the second equation of motion for the
Jackiw-Teitelboim model, which gives the vanishing torsion condition, follows
immediately from eqs. (\ref{3.14}) and (\ref{3.24}):
\begin{equation}
E^{\mu\nu} T_{\mu\nu}^a \equiv E^{\mu\nu} {\cal D}_\mu[\frac{1}{e_G}w] e_\nu^a
=0
\; . \label{3.30}
\end{equation}
Concerning the new terms induced by scalar fields, they act as an effective
energy momentum
tensor trace $\tau$. Hence, our topological model can be interpreted as a
theory
for two-dimensional gravity non-trivially coupled to matter. This has been
achieved by using a two-dimensional topological model defined through
a classical action given by eq.(\ref{2.1}). The fact that $S_{cl}$ is
constructed
from self-dual auxiliary fields allows terms such as $<E^{\mu\nu} B_{\mu\nu}
B>$ or $<E^{\mu\nu} H_\nu B_\mu>$ to be present; they induce matter
interactions
in the sense that there is no dependence on the metric at the quantum level,
as it will be demonstrated in the next Section.
Had we started from a topological action \`a la
Baulieu-Singer \cite{BS} (as in refs.\cite{CHW}-\cite{IT}), we would have faced
the problem mentioned by Chamseddine and Wyler \cite{CHW}: matter interactions
would require the introduction
of a metric in a non-trivial way (thus imposing non-covariant couplings of the
gauge field once
it has been identified with the Zweibein) or rather complicated terms.

Let us now study the equations for matter, namely
eqs.(\ref{3.16})-(\ref{3.19}). For the sake of clarity we shall distinguish
between the $\Lambda=0$ and the $\Lambda\neq 0$ cases.

\vspace{0.2in}
A)  $\Lambda=0$.

In this case, from eq.(\ref{3.19}) we have that $\psi$
is constant
\begin{equation}
\psi=\eta
\end{equation}
and, from eq.(\ref{3.18}) we can in principle determine the other components
of the $\Psi$ field in terms of the Zweibein and the other scalar field $\Phi$:
\begin{equation}
\Delta_\mu^a(w,\psi_b,\eta) - e_G \, \epsilon^{ab} \, e_\mu^c \, \psi_c \psi_b
\,
v(\Phi) =0
\; . \label{3.48}
\end{equation}

Concerning the $\Phi$ field, eq.(\ref{3.17}) implies that also
$\phi$ is constant
\begin{equation}
\phi=\lambda
\; .   \label{3.49}
\end{equation}
After some calculations, it can be shown that eq.(\ref{3.16}) reduces
to the following pair of equations
\begin{eqnarray}
(\eta^2 -1) \; [ \epsilon^{ab} \psi_a \Delta^\mu_b (f, \phi_b, \lambda)] &=& 0
 \; , \label{3.50} \\
(\eta^2 -1) \; [\psi^a \Delta^\mu_a(f, \phi_b,\lambda)] &=& 0  \label{3.51}
\; .
\end{eqnarray}

These equations have as one obvious possible solution $\eta=\pm 1$. If this
were the case, the constraints
reduce to $\psi^a\psi_a=0$ and $\phi^a \psi_a \pm \lambda =0$. If the flat
metric $\eta^{ab}$ is Euclidean, the unique solution to the former constraint
is $\psi^a \equiv 0$, but this implies, through eq.(\ref{3.48}), the vanishing
of the Zweibein. Then, $\eta \pm 1$ is not a sensible solution in Euclidean
space-time. If, on the other
hand, the flat metric $\eta^{ab}$ is Minkowskian, the first constraint has
solutions different from zero and further analysis of the complete system
is required to find explicit solutions.

For $\eta\neq\pm 1$, the matter system reduces to eq.(\ref{3.48}) and the two
equations stemming from eqs.(\ref{3.50}) and (\ref{3.51}), supplemented
with the constraints $\psi^a\psi_a+\eta^2=1$ and $\psi^a \phi_a
+\eta\lambda=0$,
coupling $\psi^a$, $\phi^a$ and $e_\mu^a$. This is a coupled non linear
system which has to be studied together with the equation
(\ref{3.28}) for the curvature scalar.

\vspace{0.2in}
B) $\Lambda \neq 0$.

In this case, we can solve $\psi^a$ in terms of $\psi$
and the Zweibein from eq.(\ref{3.19})
\begin{equation}
\psi^a =\frac{1}{2 e_G \Lambda} \epsilon^{ab} e^\mu_b \, \partial_\mu \psi
\; .
\label{psia}
\end{equation}
Using this result and the constraint (\ref{3.20}) we obtain the following
equation for $\psi$
\begin{equation}
\Box \psi + 4 e_G^2 \, (1-\psi^2) \, v(\Phi) + 4 e_G^2 \Lambda \, \psi =0
\; .
\label{3.B1}
\end{equation}
Once again, were the scalar field $\Phi$ absent we would recover the
Klein-Gordon equation in de Sitter space for the model of ref.\cite{CHW}. The
additional
term we have corresponds to a self-interaction, highly non-linear and typical
of theories with a Higgs potential.

It is still pending the study of the equations (\ref{3.16}) and (\ref{3.17}).
The analysis of the former is similar to the case $\Lambda=0$; it just appears
one extra term in each of the equations (\ref{3.50}) and (\ref{3.51})
\begin{eqnarray}
(\psi^2-1) \; [ \epsilon^{ab} \psi_a \Delta^\mu_b(f,\phi_b,\phi)] -
E^{\mu\nu} \partial_\nu \phi] &=& 0 \; , \label{3.55} \\
(\psi^2-1) \; [\psi^a \Delta^\mu_a(f,\phi_b,\phi) + \psi \partial^\mu
\phi] &=& 0 \; . \label{3.56}
\end{eqnarray}
Though $\psi^2=1$ is a solution to these equations, recalling
eq.(\ref{3.B1}) we see that it solves the complete system only if
$\Lambda$ equals
zero. Hence, we have to leave aside $\psi=\pm 1$
and study the vanishing of the brackets in eqs.(\ref{3.55}) and
(\ref{3.56}),
together with eq.(\ref{3.B1}). We arrive at the following equation
\begin{equation}
\partial_\mu \phi (1-\Lambda) -\frac{1}{2} \partial_\mu \psi =0
 \label{3.57}
\end{equation}
which distinguishes between $\Lambda$ equal or different from one. In the
former case, $\psi$ must be a constant which implies (through
eq.(\ref{psia})) that $\psi^a$ vanishes and then that $\eta$ must be equal to
$\pm 1$, leaving no solutions to the system.
In the latter case, one has to select a given manifold $M$ in order to go
further. For example, if we take $M$ to be a manifold with boundary, the
solution to (\ref{3.57}) can be written in the form
\begin{equation}
\phi= \frac{1}{2 (1-\Lambda)} \psi +(\phi_0 -\frac{\psi_0}{2(1-\Lambda)})
\end{equation}
where we have imposed $\psi \rightarrow \psi_0$ and $\phi \rightarrow \phi_0$
at the boundary.

\vspace{0.2in}
In general, the complete resolution of the full system
(\ref{3.14})-(\ref{3.19}) both in
the $\Lambda=0$
and $\Lambda \neq 0$ cases, depends on the topological structure of the
two-dimensional manifold $M$. Two different
situation
can be envisaged:\newline
\indent 1. $M$ is such that there exists a finite number of isolated
classical
solutions, that is, the ``moduli space'' ${\cal M}$ contains a finite number of
points. The dimension of ${\cal M}$ is then $d({\cal M})=0$.\newline
\indent 2. $M$ is such that the moduli space has dimension different from zero,
$d({\cal M}) \neq 0$.
\newline
We shall come back to this point, in connection with the evaluation of
topological invariants, in the next Section.

We summarize in table \ref{tabla1} what we have learnt about the equations of
motion and their solutions.

\pagestyle{empty}
{ \small  {\small
\begin{table}[h]
\centering
\begin{tabular}{|c|c|c|}    \hline
& & \\
&$\Lambda=0$
&$\Lambda \neq 0$
\\
& & \\
\hline \hline
& & \\
\small{Scalar}& & \\
\small{curvature}
&$R=e_G\tau$
&$R+2 e_G^2 \Lambda = e_G \tau$
\\
\small{equation}& & \\
& & \\
\hline
& & \\
\small{Vanishing}& & \\
\small{torsion}
&$ D_\mu^{ab} [\frac{w}{e_G}] e_\nu^b =0$
&$D_\mu^{ab} [\frac{w}{e_G}] e_\nu^b =0$
\\
\small{equation}& & \\
& & \\
\hline \hline
& & \\
&$\psi=\eta$
&$\Box \psi + 4 e_G^2 \, (1-\psi^2) \, v(\Phi) + 4 e_G^2 \Lambda \, \psi =0$
\\
& & \\
\cline{2-3}
$\Psi$  \small{field} & & \\
&\begin{tabular}{c}
$\Delta_\mu^a (f,\psi_b,\eta)=0$ \\
\end{tabular}
&$\psi^a =\frac{1}{2 e_G \Lambda} \epsilon^{ab} e^\mu_b \partial_\mu \psi$
\\
& & \\
\hline \hline
& & \\
&$\phi=\lambda$
&\begin{tabular} {ll}
$\Lambda \neq 1$:&$\phi= \frac{1}{2 (1-\Lambda)} \psi +(\phi_0
-\frac{\psi_0}{2(1-\Lambda)})$ \\
$\Lambda =1$:&\small{no solutions} \\
\end{tabular}
\\
& & \\
\cline{2-3}
$\Phi$ field & & \\
&
\begin{tabular}{ll}
$\eta=\pm 1$:&$\phi^a$ \small{to be determined} \\
&\small{from the constraint} \\
$\eta\neq\pm 1$:&$\epsilon^{ab} \psi_a \Delta^\mu_b(f,\phi^a, \lambda)
=0$ \\
&$\psi^a \Delta^\mu_a (f,\phi^a, \lambda)=0$
\\
\end{tabular}
& \begin{tabular}{rcl}
$\epsilon^{ab} \psi_a \Delta^\mu_b(f,\phi_b, \phi)-E^{\mu\nu} \partial_\nu
\phi$ &=& $0$ \\
$\psi^a \Delta^\mu_a(f,\phi_b,\phi) +\psi \partial^\mu \phi$ &=& $0$
\\
\end{tabular}
\\
& & \\
\hline \hline
& & \\
\small{Constraints}
&
\begin{tabular}{c}
$\psi^a\psi_a +\eta^2=1$ \\
$\psi^a\phi_a + \eta \lambda =0$ \\
\end{tabular}
&
\begin{tabular}{c}
$\Lambda \psi^a\psi_a +\psi^2=1$ \\
$\Lambda \psi^a\phi_a + \psi \phi =0$  \\
\end{tabular} \\
& & \\
\hline
\end{tabular}
\caption{Equations of motion and their solutions.}
\label{tabla1}
\end{table}}}

\pagestyle{plain}

\section{SYMMETRIES AND QUANTUM {\-} ACTION}
\label{section4}
\setcounter{equation}{0}

\subsection{Symmetries of the gravitational model}
\label{subsymmetries}

It is interesting to recover from topological transformation laws
(\ref{2.17})-(\ref{2.19}) the usual transformation laws of two-dimensional
gravity, viz. diffeomorphism and Lorentz transformations.

Let us start by writing the parameters $\epsilon_\mu$, $\epsilon$, $\theta$
and $\tilde\theta$ appearing in (\ref{2.17})-(\ref{2.19}) in the form
\begin{eqnarray}
\begin{array}{rcl}
\epsilon_\mu &=& \varepsilon_\mu^a P_a + \varepsilon J \; , \\
\epsilon &=& \varepsilon^a P_a + \varepsilon J \; , \\
\theta &=& \vartheta^a P_a +\vartheta J \; , \\
\tilde\theta &=& \tilde\vartheta^a P_a + \tilde\vartheta J \label{pa4}
\; .
\end{array}
\label{pa1}
\end{eqnarray}
With this, the transformation laws for $e_\mu^a$, $f_\mu$, $\phi^a$, $\phi$,
$\psi^a$ and $\psi$ can be readily recognized to be
\begin{eqnarray}
\delta e_\mu^a &=& \varepsilon_\mu^a - \partial_\mu \varepsilon^a -
e_G \, \epsilon^{ab} (\varepsilon e_{\mu b} - f_\mu \varepsilon_b) \; ,
\label{t.1} \\
\delta f_\mu  &=& \varepsilon_\mu - \partial_\mu \varepsilon + e_G \, \Lambda
\epsilon_{ab}  \, \varepsilon^a e_\mu^b \;, \label{t.2} \\
\delta \phi^a   &=& \vartheta^a +e_G \, \epsilon^{ab} \, (\phi \varepsilon_b -
\varepsilon \phi_b) \; , \label{t.3} \\
\delta   \phi &=& \vartheta - e_G \, \Lambda \epsilon_{ab} \phi^a \varepsilon^b
\; , \label{t.4} \\
\delta   \psi^a &=& \tilde  \vartheta^a + e_G \, \epsilon^{ab} (\psi
\varepsilon_b - \varepsilon \psi_b)  \; , \label{t.5} \\
\delta   \psi &=& \tilde  \vartheta - e_G \Lambda \epsilon_{ab} \, \psi^a
\varepsilon^b
\; .   \label{t.6}
\end{eqnarray}

Our first purpose is to compare these transformations with diffeomorphism
transformations $\delta_D$
\begin{eqnarray}
\delta_D e_\mu^a &=& v^\alpha (\partial_\alpha e_\mu^a - \partial_\mu
e_\alpha^a) +\partial_\mu (v^\alpha e_\alpha^a) \; , \label{d.1}\\
\delta_D f_\mu &=& v^\alpha (\partial_\alpha f_\mu - \partial_\mu
f_\alpha) +\partial_\mu (v^\alpha f_\alpha) \; , \label{d.2} \\
\delta_D \Phi &=& v^\alpha \partial_\alpha \Phi  \; , \label{d.3}\\
\delta_D \Psi &=& v^\alpha \partial_\alpha \Psi \; , \label{d.4}
\end{eqnarray}
where $v^\alpha$ is the local parameter transforming $x^\alpha$,
$\delta_D x^\alpha=v^\alpha$.
In order to find a connection between topological and diffeomorphism
transformations, let us consider the following subset of parameters
\begin{eqnarray}
\begin{array}{rcl}
\varepsilon^a &=& -\frac{1}{e_G} v^\alpha e_\alpha^a \; , \\
\varepsilon &=& -\frac{1}{e_G} v^\alpha f_\alpha \; , \\
\varepsilon_\mu^a &=& v^\alpha E_{\alpha\mu} \psi^a v(\Phi)
 \; , \\
\varepsilon_\mu &=& v^\alpha E_{\alpha\mu} \psi v(\Phi) \; .
\end{array}
\label{p.1}
\end{eqnarray}
With this we find, from (\ref{t.1}) and (\ref{d.1}),
\begin{equation}
\delta e_\mu^a -\delta_D e_\mu^a =  \frac{1}{e_G} v^\alpha [{\cal D}_\mu^{ab}
{[}f{]} e_{\alpha b} -
{\cal D}_\alpha^{ab} {[}f{]} e_{\mu b}- e_G E_{\mu\alpha} \psi^a v(\Phi) ]
\end{equation}
or, using the equation of motion (\ref{3.14}) \footnote{We represent the
use of the equations of motion by $|_{o.s}$.}
\begin{equation}
\delta e_\mu^a -\delta_D e_\mu^a|_{o.s.} = -\frac{1}{e_G} v^\alpha
h_{\alpha\mu}^a
\; .
\end{equation}
Concerning $f_\mu$, a similar procedure shows that the difference between
large and diffeomorphism transformations is, using equation (\ref{3.15}),
\begin{equation}
\delta f_\mu - \delta_D f_\mu |_{o.s.}= -\frac{1}{e_G} v^\alpha h_{\alpha\mu}
\; .
\end{equation}
With respect to the scalar field $\Psi$, once the parameters $\epsilon$ and
$\epsilon_\mu$ have
been fixed, it is simple to prove from eqs.(\ref{t.5})
and (\ref{d.4}) the following identity
\begin{equation}
\delta\psi^a-\delta_D \psi^a = \tilde \vartheta^a - \frac{1}{e_G} v^\alpha
\Delta_\alpha^a(f,\psi_b,\psi)
\end{equation}
then, choosing $\tilde\vartheta^a=0$ and using the equation of motion
(\ref{3.18}) we have
\begin{equation}
\delta\psi^a-\delta_D \psi^a |_{o.s.}= -\frac{1}{e_G} v^\alpha \tilde
h_\alpha^a
\;.
\end{equation}
The analysis for $\psi$ is analogous: choosing $\tilde\vartheta=0$ and
using eq.(\ref{3.19})
\begin{equation}
\delta\psi-\delta_D \psi|_{o.s.} = -\frac{1}{e_G} v^\alpha \tilde h_\alpha
\; .
\end{equation}
Finally, the difference between variations of the components of the field
$\Phi$ are
\begin{eqnarray}
\delta \phi^a-\delta_D \phi^a &=& \vartheta^a - \frac{1}{e_G} v^\alpha
\Delta_\alpha^a(f,\phi_b,\phi) \; , \\
\delta \phi-\delta_D \phi &=& \vartheta - \frac{1}{e_G} v^\alpha
(\partial_\alpha \phi
+ e_G \Lambda \, \epsilon_{ab} \phi^a e_\alpha^b)
\; .
\end{eqnarray}
Now, we can select $\vartheta^a$ and $\vartheta$ in the following way
\begin{eqnarray}
\vartheta^a &=& \frac{1}{e_G} v^\alpha \Delta_\alpha^a(f,\phi_b,\phi) \; , \\
\vartheta &=& \frac{1}{e_G} v^\alpha (\partial_\alpha \phi + e_G \Lambda
\, \epsilon_{ab} \phi^a e_\alpha^b) \; ,
\end{eqnarray}
which implies
\begin{eqnarray}
\delta \phi^a - \delta_D \phi^a &=& 0 \; , \\
\delta \phi - \delta_D \phi &=& 0 \; .
\end{eqnarray}
Similarly, we can show that the difference between a topological transformation
((\ref{2.20})-(\ref{2.22})) and a diffeomorphism transformation for each
auxiliary field is proportional to the corresponding auxiliary field.

In summary, working in the gauge in which all auxiliary fields vanish, $\delta
H_{\mu\nu}=\delta_D H_{\mu\nu}=0$,
$\delta H_\mu=\delta_D H_\mu = 0$, $\delta \tilde H_\mu=\delta_D \tilde
H_\mu=0$, we have
\begin{eqnarray}
\begin{array}{rcl}
\delta e_\mu^a - \delta_D e_\mu^a |_{o.s} &=& 0 \; , \\
\delta f_\mu - \delta_D f_\mu |_{o.s} &=& 0 \; , \label{dif2}\\
\delta \Phi - \delta_D \Phi |_{o.s} &=& 0 \; , \label{dif3}\\
\delta \Psi - \delta_D \Psi|_{o.s} &=& 0
\;.
\end{array}
\label{dif1}
\end{eqnarray}

Concerning Lorentz transformations $\delta_L$, again an appropriate choice
of parameters allows their identification with transformations
(\ref{t.1})-({\ref{t.6}). Indeed, if we choose
\begin{equation}
\varepsilon_\mu^a = \varepsilon_\mu = \varepsilon^a = \vartheta^a = \vartheta=
\tilde\vartheta^a = \tilde\vartheta =0 \label{p.5}
\end{equation}
and
\begin{equation}
\varepsilon = - \frac{\kappa}{e_G}
\label{p.6}
\end{equation}
we have
\begin{eqnarray}
\begin{array}{rcl}
\delta e_\mu^a &=& \kappa \epsilon^{ab} e_{\mu b}
\;, \label{L.0}\\
\delta f_\mu   &=& \partial_\mu \kappa \; ,
\label{L.1}\\
\delta \phi^a  &=& \kappa \epsilon^{ab} \phi_b  \; ,
\label{L.2}\\
\delta \psi^a  &=& \kappa \epsilon^{ab} \psi_b \; ,
\label{L.3}\\
\delta \phi    &=& 0   \; ,
\label{L.4}\\
\delta \psi    &=& 0
\label{L.5}
\; .
\end{array}
\end{eqnarray}
The right hand side of eqs.(\ref{L.5}) precisely corresponds to Lorentz
transformations $\delta_L$ with parameter $\kappa$ and then,
\begin{eqnarray}
\begin{array}{rcl}
\delta e_\mu^a &=& \delta_L e_\mu^a \; , \\
\delta f_\mu &=& \delta_L f_\mu \; , \\
\delta \Phi &=& \delta_L \Phi \; , \\
\delta \Psi &=& \delta_L \Psi
\label{lorentz}
\end{array}
\end{eqnarray}
and $H_{\mu\nu} = \delta_L H_{\mu\nu}=0$,
$\delta H_\mu=\delta_L H_\mu =0$, $\delta \tilde H_\mu =\delta_L \tilde
H_\mu=0$ in the gauge in which all auxiliary fields vanish.

We then see from eqs.(\ref{dif1}) and
(\ref{lorentz}) that, as expected, the topological model defined from the
classical
action (\ref{2.1}) can be used as a model for two-dimensional gravity with
its topological transformations interpreted as diffeomorphism and Lorentz
transformations. In order to make such an identification we have restricted
the parameter space to a subspace satisfying (\ref{p.1}),
(\ref{p.5}) and (\ref{p.6}) relations. In this sense, the whole topological
invariance is larger than the usual invariances for gravity.

\subsection{The Quantum Action}

As explained in Section \ref{Section2}, because of the large topological
symmetry (eqs. \linebreak
(\ref{2.17})-(\ref{2.22})) of the classical action (\ref{2.1}),
one has to proceed to a careful BRST
quantization in which ghosts and ghosts for ghosts appear through the process
of gauge fixing. We shall skip the details (given in ref.\cite{CLS1} for
the gauge theory and sketched in Section \ref{Section2}) and just quote the
result for the quantum action
\begin{eqnarray}
S_q [M] &=&  \int_M d^2x \sqrt g <
B_{\mu\nu} D^{\mu\nu} -\frac{1}{4} D_{\mu\nu} D^{\mu\nu}
+ B_\mu D^\mu + \nonumber \\
& &
\;\;\;\;\;\;\;\;\;\;\;\;
\tilde B_\mu \tilde D^\mu > + S_{gf}[M] + S_{gh}[M]
\; .
\label{S_q}
\end{eqnarray}
The explicit expression for the classical part of $S_q$ in gravity language is
straightforwardly
obtained calculating the adequate inner product and
expressing $B_{\mu\nu}$, $B_\mu$ and $\tilde B_\mu$ components as in
eqs.(\ref{b1})-(\ref{b2}). In (\ref{S_q}) auxiliary
fields $H_{\mu\nu}$, $H_\mu$ and $\tilde  H_\mu$ have been traded for
Lagrange multipliers $D_{\mu\nu}=(d_{\mu\nu}^a, d_{\mu\nu})$,
$D_\mu=(d_\mu^a,d_\mu)$ and $\tilde  D_\mu=(\tilde d_\mu^a,\tilde d_\mu)$. Of
course, the equations of motion arising from this classical part coincide
with those gotten from eq.({\ref{2.1}) in the $H_{\mu\nu}=0$, $H_\mu=0$ and
$\tilde H_\mu=0$ gauge, and also the metric and coupling constant independence
is maintained. From the explicit expression of $S_q$ one
also sees that $e_G^2$ can be identified with Newton's
gravitational constant.

Concerning the gauge fixing action $S_{gf}$, it cannot be expressed in a
covariant way and the introduction of a metric is unavoidable. The metric
$g_{\alpha\beta}$ on $M$
selected to incorporate matter couplings is here again used; evidently,
physical results should be independent of this choice. A particularly
advantageous gauge is the Landau gauge. In order to
appropriately introduce it we define a covariant
derivative ${\cal D}_\alpha [e_{cl}, w_{cl}]$ which acts
on a vector $C_\beta$ taking values in the algebra of the gauge group
in the following way
\begin{eqnarray}
{\cal D}_\alpha [e_{cl}, w_{cl}]
C_\beta &=& ( \partial_\alpha c_\beta^a -
\Gamma^\sigma_{\alpha\beta}[g] \, c_\sigma^a - e_G \, \epsilon^{ab}
{w_{cl}}_\alpha c_{\beta b} ) \, P_a + \\
& & ( \partial_\alpha
c_\beta - \Gamma^\sigma_{\alpha\beta}[g] \, c_\sigma + e_G \Lambda \,
\epsilon_{bc} {e_{cl}}^b_\alpha c_\beta^c ) \, J
\; .
\end{eqnarray}
Here, we have used the gravitational covariant derivative plus a term
containing background Zweibein and spin connection fields $e_{cl}$ and
$w_{cl}$ which are solutions to the equations of motion (these last have
been introduced to handle with zero mode problems). With this notation
the Landau gauge condition reads
\begin{eqnarray}
{\cal D}^\alpha [e_{cl}, w_{cl}] (e_\beta, f_\beta) &=& 0 \\
{\cal D}^\alpha [e_{cl}, w_{cl}] \chi_\beta &=& 0
\end{eqnarray}
and $S_{gf}$ is
\begin{eqnarray}
S_{gf}[M] &=&  \int_M d^2x \sqrt g  <
Y {\cal D}_\mu[e_{cl}, w_{cl}] (e^\mu, f^\mu) + \nonumber \\
& & \hspace{0.6in} \tilde Y {\cal D}_\mu[e_{cl}, w_{cl}] \chi^\mu  >
\end{eqnarray}
where $Y=(y^a,y)$ and $\tilde Y=(\tilde y^a, \tilde y)$ are Lagrange
multipliers enforcing the gauge conditions.
The corresponding ghost action takes the form
\begin{eqnarray}
S_{gh}[M] &=&  \int_M d^2x \sqrt g <
X(E^{\mu\nu} D_\mu [e, f] \chi_\nu
- 2 e_G\, \Psi <\Phi\rho> -
\nonumber \\
& &
\;\;\;\;\;\;\;\;\;\;
e_G <(\Phi^2 - \Phi^2_0) \tilde\rho>) +
\frac{1}{4} e_G \, [X,X] \sigma + X_\mu^+
(-e_G [\Phi,\chi^\mu] +
\nonumber \\
& &
\;\;\;\;\;\;\;\;\;\;
D^\mu[e, f] \rho -
E^{\mu\nu} [\Psi,[\Phi,\chi_\nu]] + E^{\mu\nu} [ \Psi, D_\nu [e,f]\rho] -
\nonumber \\
& &
\;\;\;\;\;\;\;\;\;\
e_G \, E^{\mu\nu} [ D_\nu [e, f] \Phi,
\tilde\rho]) +
\frac{1}{4} e_G \, [ X_\mu, X^\mu] \sigma  +
\nonumber \\
& &
\;\;\;\;\;\;\;\;\;\
\tilde X_\mu^+
(-e_G [\Psi,\chi^\mu]+
D^\mu[e, f]
\tilde\rho) +
\frac{1}{4} e_G \, [\tilde X_\mu,\tilde X^\mu] \sigma +
\nonumber \\
& &
\;\;\;\;\;\;\;\;\;\;
(-{\cal D}_\mu[e_{cl}, w_{cl}] \overline
C +e_G [\chi_\mu, \overline \sigma]) (\chi^\mu - D^\mu[e, f] C) +
\nonumber \\
& &
\;\;\;\;\;\;\;\;\;\;
\overline \sigma {\cal D}_\mu[e_{cl}, w_{cl}]
(D^\mu[e,f]\sigma + e_G[C, \chi^\mu] )\; >
\; .
\label{sgh}
\end{eqnarray}
Fields ${\cal C}=(\chi_\mu, C, \rho, \tilde\rho, \sigma)$ with ghost
numbers $(1,1,1,1,2)$ are the ghosts associated with each of the symmetries of
the classical action. To be more precise, they are related as follows
\begin{eqnarray}
\epsilon_\mu &\rightarrow& \chi_\mu \; , \label{chi} \\
\epsilon &\rightarrow& C \; \\
\theta &\rightarrow& \rho \; \\
\tilde\theta &\rightarrow& \tilde\rho \; \\
\xi &\rightarrow& \sigma \;.
\end{eqnarray}
The corresponding antighosts are written as
$\overline {\cal C} = (\overline\sigma, \overline C, X_\mu, \tilde X_\mu, X)
$ with
ghost numbers $(-2,-1,-1,-1,-1)$. $X_\mu$, $\tilde X_\mu$ and $X$ are
self-dual fields in the sense of eqs.(\ref{2.3}) and (\ref{2.2}), respectively.
The covariant derivative $D_\mu[e,f]$
has been introduced in eq.(\ref{cov}).

The partition function for our
model, when written in gravity language is, then,
\begin{equation}
{\cal Z}[M] = \int {\cal D}{\mbox{fields}} \;\; e^{-S_q[M]}
\; .
\end{equation}
The fields of the theory and their corresponding ghost numbers
are summarized in table \ref{tabla2}.

Given the topological invariance of the action $S_{cl}$
(eqs.(\ref{t.1})-(\ref{t.6})), it is easy to find the associated BRST
commutators
(\ref{2.24}) for gravity and matter fields
\begin{eqnarray}
\begin{array}{rcl}
\{Q, e_\mu^a \} &=& \chi_\mu^a - \partial_\mu c^a - e_G \,
\epsilon^{ab} (c e_{\mu b} - f_\mu c_b) \; , \\
\{Q, f_\mu \}   &=& \chi_\mu - \partial_\mu c + e_G \Lambda \,
\epsilon_{ab}  \, c^a e_\mu^b \; ,   \\
\{Q, \Phi\}  &=& \rho - e_G [\Phi, C] \; , \\
\{Q,\Psi \}   &=& \tilde  \rho - e_G [\Psi, C] \; ,
\end{array}
\label{BRSTcampos}
\end{eqnarray}
for ghosts and antighosts
\begin{eqnarray}
\begin{array}{rclrcl}
\{Q, \chi_\mu \} &=& - D_\mu [e, f]
\sigma + e_G [C,\chi_\mu]
\; ,
& \{Q, C \} &=& - (\sigma +\frac{1}{2} e_G [C,C]) \; ,
\nonumber \\
\{Q, \rho \} &=& -e_G \, ([\Phi,\sigma] + [C,\rho]) \; ,
& \{Q, \sigma \} &=& e_G \, [\sigma,C] \;,
\nonumber \\
\{Q, \tilde \rho\} &=& -e_G \, ([\Psi,\sigma] + [C,\tilde \rho]) \; ,
&
\{Q, \overline C\} &=& Y \; ,
\nonumber \\
\{Q, \overline \sigma\} &=& \tilde Y \; ,
\nonumber \\
\end{array}
\end{eqnarray}
and for Lagrange multipliers
\begin{eqnarray}
\begin{array}{rcl}
\{Q, Y\} &=& 0 \; , \nonumber \\
\{Q, \tilde Y \} &=& 0 \; , \nonumber \\
\{Q, X \} &=& \frac{1}{2} E^{\mu\nu} D_{\mu\nu} - e_G [X ,C] \; ,
\nonumber \\
\{Q, X_\mu \} &=& D_\mu -  e_G [X_\mu,C] \; ,
\nonumber \\
\{Q, \tilde X_\mu \} &=& \tilde D_\mu - e_G [\tilde X, C] \; ,
\nonumber \\
\{Q, D_{\mu\nu} \} &=& e_G\, ([D_{\mu\nu},C] + E_{\mu\nu} [X,\sigma]) \;,
\nonumber \\
\{Q, D_\mu \} &=& e_G\, ([D_\mu,C] +[X_\mu, \sigma]) \; ,
\nonumber \\
\{Q, \tilde D_\mu \} &=& e_G \, ([\tilde D_\mu, C] +
[\tilde X_\mu, \sigma]) \; .
\nonumber \\
\end{array}
\end{eqnarray}
It is straightforward but tedious to corroborate the BRST invariance of $S_q$.
Moreover, it can be also proved that, as announced in the previous Section,
\begin{equation}
S_q = \{Q,V \}
\label{V}
\end{equation}
where the functional $V$ is
\begin{eqnarray}
V&=& \int_M d^2x \sqrt g < \frac{1}{4} X E^{\mu\nu}
D_{\mu\nu} -X E^{\mu\nu} B_{\mu\nu} +
\frac{1}{4} X^\mu D_\mu - \nonumber \\
& &
\;\;\;\;\;\;\;\;\;\;\;\;\;\;\;
X^\mu B_\mu + \frac{1}{4} \tilde X^\mu \tilde D_\mu -
\tilde X^\mu \tilde B_\mu -\nonumber \\
& &
\;\;\;\;\;\;\;\;\;\;\;\;\;\;\;
\overline C {\cal D}_\mu [e_{cl}, w_{cl}] (e^\mu, f^\mu)
-\overline\sigma {\cal D}_\mu [e_{cl}, w_{cl}] \chi^\mu  >
\; .
\end{eqnarray}
This property guarantees that ${\cal Z}[M]$ only depends on the topology
of $M$ and not on the choice of the selected metric. In fact,
\begin{equation}
\frac{\delta S_q}{\delta g^{\mu\nu}} = \{ Q, \frac{\delta V}{\delta g^{\mu\nu}}
\}
\; ,
\end{equation}
which ensures that the metric dependence of the quantum action is trivial
in the sense that its variation with respect to the metric gives a BRST
commutator which has no effect at the physical level.
More precisely, a possible dependence of the partition function measure
on the metric must be taken into account to finally establish the
independence of ${\cal Z}$ on the metric. This has been done in
ref.\cite{medida} for Witten type TQFTs and it has been there confirmed that,
or this kind of theories, ${\cal Z}$ is indeed metric independent.
Furthermore, (\ref{V}) implies the independence of ${\cal Z}[M]$
on the gauge coupling constant $e_G$.

\pagestyle{empty}
%
{ \small
\begin{table}[h]
\centering
\begin{tabular}{|c|c|c|}    \hline
& & \\
Field & & Ghost number \\
& & \\
\hline\hline
& & \\
Zweibein & $e_\mu^a$ & $0$ \\
(related to the spin connection) & $f_\mu$ & $0$ \\
Scalar field & $\Phi$ & $0$ \\
Scalar field & $\Psi$ & $0$ \\
& & \\
\hline
& & \\
Lagrange multipliers &
\begin{tabular}{c}
$D_{\mu\nu}$\\
$D_{\mu}$ \\
$\tilde D_\mu$\\
$Y$ \\
$\tilde Y$ \\
\end{tabular}
&
\begin{tabular}{c}
$0$ \\
$0$ \\
$0$ \\
$0$ \\
$0$ \\
\end{tabular}
\\
& & \\
\hline
& & \\
Ghost fields &
\begin{tabular}{c}
$\chi_\mu$ \\
$C$ \\
$\rho$ \\
$\tilde \rho$ \\
$\sigma$ \\
\end{tabular}
&
\begin{tabular}{c}
$1$ \\
$1$ \\
$1$ \\
$1$ \\
$2$ \\
\end{tabular}
\\
& & \\
\hline
& & \\
Antighost fields &
\begin{tabular}{c}
$\overline \sigma$ \\
$\overline C$  \\
$X_\mu$ \\
$\tilde X_\mu$ \\
$X$ \\
\end{tabular}
&
\begin{tabular}{c}
$-2$ \\
$-1$ \\
$-1$ \\
$-1$ \\
$-1$ \\
\end{tabular}
\\
& & \\
\hline
\end{tabular}
\caption{Fields of the theory, ghost numbers and Grassmann parities.}
\label{tabla2}
\end{table}
}

\pagestyle{plain}

\section{TOPOLOGICAL INVARIANTS}
\setcounter{equation}{0}
\label{Section4}

In view of the independence of the partition function on the metric signaled
above, the simplest topological invariant to be considered is,
precisely, the partition function ${\cal Z}[M]$.

In order to clarify our derivation of topological invariants,
we shall again first consider the simplified
action $S_{cl}^1$, eq.(\ref{S1}).
It can be easily shown that the zero mode equation associated with the ghost
field $\chi$ appearing in $S_q^1$ coincides with the equation describing the
moduli space for Bogomol'nyi solutions. Indeed, given a solution $\Phi_{cl}$
to Bogomol'nyi equations,
\begin{equation}
B[\Phi_{cl}]=0
\; ,
\end{equation}
a nearby configuration $\Phi_{cl} + \delta \Phi_{cl}$ will also be a solution
provided
\begin{equation}
\frac{\delta B}{\delta \Phi} |_{\Phi_{cl}} \delta \Phi_{cl} =0
\; .
\label{mod}
\end{equation}
Since the ghost action in $S_q^1$ is
\begin{equation}
S_{gh}^1= \int_M d^2x \sqrt g < X \frac{\delta B}{\delta
\Phi} \chi >
\; ,
\end{equation}
the equation of motion for $X$, giving the zero mode equation for $\chi$,
coincides with eq.(\ref{mod}) for $\delta \Phi_{cl}$ when $\Phi=\Phi_{cl}$.
(For simplicity we shall suppose that $X$ has no zero modes.)

As for solutions to eq.(\ref{mod}), there are two possibilities; either
no non-trivial solution exists or there are solutions which span the moduli
space; $d({\cal M})$ is equal or different from zero, respectively.

Concerning the case $d({\cal M})=0$, ${\cal Z}[M]$ can be exactly evaluated,
a basic property of topological models, related to the $Q$-symmetry of $S_q$.
Indeed, ${\cal Z}[M]$ is independent of $e_G$ and then it can be
computed in the $e_G$ going to zero limit where the path integral is dominated
by configurations $(\Phi, \chi, X)=(\Phi_{cl}^i, 0, 0)$, with $i=1,2,...n$
labelling isolated Bogomol'nyi solutions. Calling $\varphi$ the fluctuations
around $\Phi=\Phi_{cl}^i$ we have
\begin{eqnarray}
{\cal Z}[M] &=& \sum_{i=1}^n \int {\cal D} \varphi \;  {\cal D} \chi \;  {\cal
D} X \;  \exp[- \int_M d^2 x \sqrt g < \varphi \frac{\delta
B}{\delta \Phi}|_{\Phi_{cl}^i} \frac{\delta B}{\delta \Phi}|_{\Phi_{cl}^i}
\varphi
+ \nonumber \\
& & \hspace{0.6in} X \frac{\delta B}{\delta \Phi}|_{\Phi_{cl}^i} \chi >]
\end{eqnarray}
or
\begin{eqnarray}
{\cal Z}[M] &=& \sum_{i=1}^n \frac { \mbox{Pfaff}(\frac{\delta B}{\delta
\Phi}|_{\Phi_{cl}^i})}
{\sqrt{\det(\frac{\delta B}{\delta \Phi}|_{\Phi_{cl}^i} \frac{\delta B}{\delta
\Phi}|_{\Phi_{cl}^i}) }} \; , \nonumber \\
{\cal Z}[M]&=& \sum_{i=1}^n (-1)^{n_i}
\label{z}
\end{eqnarray}
where $n_i=0,1$ according to the way one determines the sign of the Pfaffian
(see ref.\cite{W1}). Since in topological theories ${\cal Z}[M]$ is metric
independent, the right hand side of eq.(\ref{z}) gives the explicit way
of computing a topological invariant.

The derivation we have presented for this simple example can be
straightforwardly extended to the model of interest with classical action
(\ref{2.1}). Simply, in view of the symmetry (\ref{2.17})-(\ref{2.22}), the
gauge fermion $F$ has been taken as
\begin{eqnarray}
F&=&  \int_M d^2x \sqrt g <
XH + X_\mu H^\mu + \tilde X_\mu \tilde H^\mu + \overline C {\cal D}_\mu
[e_{cl}, w_{cl}] (e^\mu, f^\mu) + \nonumber \\
& & \hspace{0.4in} \overline \sigma {\cal D}_\mu [\tilde
e, w_{cl}] \chi^\mu >
\; ,
\end{eqnarray}
so that the quantum action, when written in terms of gravitational fields,
is given by eq.(\ref{S_q}). Again, the bosonic and fermionic contributions to
${\cal Z}[M]$
cancel up to a sign around each classical solution. These signs
have to be computed from the quantum action for our gravity model,
eq.(\ref{S_q}). In order to do so, one first performs an expansion around
the classical solutions discussed in Section \ref{Section3} up to quadratic
terms and then computes bosonic an fermionic determinants once an
assignment
for the Pfaffian sign is adopted. Each $n_i$ can then be determined
and one can again conclude that ${\cal Z}[M]$ takes the form
\begin{equation}
{\cal Z}[M]= \sum_{i=1}^n (-1)^{n_i}
\; ,
\end{equation}
and is a topological invariant in the $d({\cal M})=0$ case.

Let us now discuss the evaluation of topological invariants in the $d({\cal
M}) \neq 0$ case. In this case, the Pfaffian vanishes and, as explained in
ref.\cite{W1}, topological
invariants have to be computed from
vacuum expectation values of BRST invariant and metric
independent functionals containing a product of an appropriate number of fields
so as to absorb zero modes. In ref.\cite{CLS1} the construction of such
invariants was discussed for the gauge theory defined by action (\ref{2.1}).
One starts by constructing functionals $W_k$ satisfying
\begin{eqnarray}
\begin{array}{rcl}
0&=& \{Q, W_0\} \; , \\
dW_0 &=& \{Q, W_1\} \; ,  \\
dW_1 &=& \{Q, W_2 \} \; ,  \\
dW_2 &=& 0 \; .
\end{array}
\end{eqnarray}
and using the notation of Section \ref{Section2} one easily finds
\begin{eqnarray}
\begin{array}{rcl}
W_0 &=& \frac{1}{2} < \sigma^2 > \; , \\
W_1 &=& < \sigma \chi_\mu > dx^\mu  \; , \\
W_2 &=& < \sigma F_{\mu\nu} >dx^\mu \wedge dx^\nu  \; .
\end{array}
\end{eqnarray}
These functionals have ghost number $4-k$. Given a moduli dimension
$d({\cal M}) \neq 0$, a non-trivial topological invariant takes the form
\begin{equation}
{\cal Z}(\gamma_1,...,\gamma_r) = \int {\cal D}{\mbox{fields}} \;\;\;
\prod_{i=1}^r I^{(\gamma_i)} \;\;\;  e^{-S_q [M]}
\; ,
\label{VEV}
\end{equation}
with $\gamma_1, ..., \gamma_r$ homology cycles of dimension $k_1, ..., k_r$
such that
\begin{equation}
\sum_{i=1}^r (4-k_i) = d({\cal M})
\end{equation}
and $I^{(\gamma_i)}$ defined as
\begin{equation}
I^{(\gamma_i)}= \int_{{\gamma}_{k_i}} W_{k_i}
\; .
\end{equation}

In order to obtain explicit formul\ae $\;$ for topological invariants, computed
as vacuum expectation values ($vev$'s) in the form (\ref{VEV}), one proceeds as
follows. As in the partition function case, the lowest order in the $e_G^2$
expansion gives the exact result for the path integral defining the $vev$,
then, the dynamical fields can be replaced by their classical configurations
solving the Bogomol'nyi equations. In the present case, the only dynamical
field appearing in $W_k$'s is the gauge field $A_\mu$ which is replaced
by $A_\mu^{cl}$. The ghost $\chi_\mu$ appearing in $W_1$, whose zero
modes probe the moduli space (together with $\rho$ and $\tilde \rho$ zero
modes), have to be replaced by its zero mode configuration $\chi_\mu^0$.
Concerning the ghost for ghost
$\sigma$, one has to perform the corresponding integration. For example,
the $vev$ of $\sigma^A$ ($\sigma= \sigma^A T_A$) is computed as follows:
\begin{eqnarray}
<\sigma^A>_{vev}&=&
\int {\cal D} \sigma \; {\cal D} \overline \sigma
\; \sigma^A(x)  \exp[- \int_M d^2y \sqrt g < \overline
\sigma D_\mu D^\mu \sigma + \nonumber \\
& & \;\;\;\;\; [\chi_\mu^0, \chi^{\mu 0}] \overline \sigma
+ ...>]
\; .
\end{eqnarray}
The dots in the exponential represent irrelevant terms to lowest order in
$e_G^2$. Expanding the second term and performing the integration over $\sigma$
and $\overline \sigma$, one has
\begin{equation}
<\sigma^A>_{vev}= \int_M d^2y \sqrt g < [\chi_\mu^0(y), \chi^{\mu
0}(y)] T_B > \Delta^{AB} (y-x)
\; ,
\label{sigma}
\end{equation}
where
\begin{equation}
(D_\mu D^\mu \Delta)^{AB} (z) = \delta^{AB} \delta (z)
\; .
\end{equation}
Replacing $\sigma$ by $<\sigma>_{vev}$ whenever it appears in
$I^{(\gamma_i)}$, one obtains the following expressions for $I^{(\gamma_i)}$'s
\begin{eqnarray}
I^{(\gamma_0)} &=& \int_{\gamma_0} < <\sigma>_{vev}^2 > \; ,
\label{gamma0} \\
I^{(\gamma_1)} &=& \int_{\gamma_1} < <\sigma>_{vev} \chi_\mu^0 >
dx^\mu \; ,
\label{gamma1} \\
I^{(\gamma_2)} &=& \int_{\gamma_2} < <\sigma>_{vev} F_{\mu\nu}^{cl}
> dx^\mu \wedge dx^\nu
\label{gamma3}
\; .
\end{eqnarray}
Of course, to go further into the evaluation of topological invariants one
has to know the structure of the moduli space, the explicit form of
$A_\mu^{cl}$, $\chi_\mu^0$, etc.

We just conclude by writing the results presented above in terms of the
fields appearing in our gravity model. The $vev$ of $\sigma$ is still given
by eq.(\ref{sigma}) with $\Delta^{AB}(z)$ satisfying
\begin{equation}
({\cal D}_\mu [e_{cl}, w_{cl}] {\cal D}^\mu [e,f] \Delta)^{AB} (z) =
\delta^{ab} \delta (z)
\; .
\label{delta}
\end{equation}
Then, $I^{(\gamma_0)}$ and $I^{(\gamma_1)}$ are computed from
eqs.(\ref{gamma0}),
(\ref{gamma1}) and (\ref{delta}) with $\chi_\mu^0$ the zero modes of the
fermionic operator in (\ref{S_q}). Concerning $I^{(\gamma_2)}$ note that
\begin{equation}
F_{\mu\nu}^{cl} dx^\mu \wedge dx^\nu = e_G \Psi^{cl} \; v(\Phi^{cl}) \; d^2x
\; ,
\end{equation}
through the use of Bogomol'nyi equation (\ref{2.7}) and then,
\begin{equation}
I^{(\gamma_2)} = e_G \int_{\gamma_2} d^2x < < \sigma>_{vev} \Psi^{cl} >
v(\Phi^{cl})
\; .
\end{equation}

\section{SUMMARY AND DISCUSSION}
\setcounter{equation}{0}

In this work, we have succeeded in constructing a two-dimensional model
for the gravitational field with a non-trivial coupling to matter. This
has been achieved starting from the topological gauge model presented in
ref.\cite{CLS1} and interpreting the gauge fields as a Zweibein and (effective)
connection fields. In this way, the original TQFT has been expressed in
geometrical terms so that its classical equations of motion become
gravitational field equations coupled to matter (see table \ref{tabla1}).
The basic property of (Witten type) TQFTs, i.e. the fact that $S_q=\{Q,
V\}$ has been fundamental to get a gravitational model with matter coupling.
Indeed, since \( \delta S_q / \delta g^{\mu\nu} = \{Q, \lambda_{\mu\nu}\}
\),
the quantum theory does not depend on the background metric used to introduce
matter couplings and to fix the gauge. The same property ensures the model
independence on all of the parameters, in particular on $\phi_0$, the minimum
of the Higgs potential. Thus the small $e_G$ expansion performed to calculate
expectation values of interest is, in this case exact and, furthermore,
the model is scale invariant.

It is interesting to point that, if all scalar fields are put to zero (i.e.
matter is absent) our equations of motion become those of the Jackiw-Teitelboim
model for two-dimensional gravity \cite{JT}. If only one scalar field (that
appearing
with a symmetry breaking potential) is set to zero, then the model becomes
that constructed by Chamseddine and Wyler \cite{CHW}. To be more precise, the
classical
equations of our model coincide with those of ref.\cite{CHW} when $\Phi$
is absent. At the quantum level, Chamseddine and Wyler quantized a topological
theory \`a la Baulieu-Singer \cite{BS}, starting from a classical action which
is
a topological invariant while we proceeded to quantization \`a la
Labastida-Pernici \cite{LP} starting from a quantum action where Bogomol'nyi
equations play a central r\^ole.

We have explicitely shown how the large symmetry, characteristic of topological
theories, corresponds to diffeomorphism and local Lorentz symmetries in
a certain subspace of transformation parameter space. Thus, as expected,
the basic gravitational symmetries are incorporated in our model.

As stated above, the exact quantum description of our model
can be made in the limit of small gauge coupling constant (which can be
here interpreted as Newton's gravitational constant). In particular,
the partition function can be computed exactly by performing a semiclassical
expansion, this leading to an explicit expression for a topological invariant
(when the moduli space dimension is zero). Other topological invariants
have been discussed by exploiting the BRST invariance of the gauge theory.

Our results extend those of refs.\cite{CHW}-\cite{IT},
in which topological theories of pure gravitational fields in two dimensions
have been constructed, to a gravity-matter theory. In all of these models,
the large topological symmetry of the action reduces the space of states
to a finite dimensional one. It would be worthwhile to investigate this
issue following, for example, Horowitz approach to the computation of state
functions for TQFTs \cite{Horowitz}, to probe whether there exists a unique
solution as
it is the case in several cases. Finally, it should be stressed that if one
takes
our model as a toy model for gravity, the large topological symmetry should
be broken. These and related problems should be studied more thoroughtfully.

$\;$\newline
{\bf Acknowledgements:}

One of the authors (L.F.C.) would like to thank Professor Abdus Salam,
the International Atomic Energy Agency and UNESCO for hospitality at the
International Centre for Theoretical Physics, Trieste.

\newpage
\pagestyle{empty}
\end{document}